%% file: main.tex
\documentclass[
twocolumn,
review=True
]{ceurart}

\usepackage{listings}
\input{datalabmacros} %
\mkclean %

\begin{document}

\copyrightyear{2023}
\copyrightclause{Copyright for this paper by its authors.
  Use permitted under Creative Commons License Attribution 4.0
  International (CC BY 4.0).}

\conference{VLDB 2023 PhD Workshop, co-located with the 49th International Conference on Very Large Data Bases (VLDB 2023), August 28, 2023, Vancouver, Canada}

\title{Discovering Dichotomies for Problems in Database Theory}

\author{Neha Makhija}[%
orcid=0000-0003-0221-6836,
email=makhija.n@northeastern.edu,
url=nehamakhija.github.io,
]
\address{Supervised by Wolfgang Gatterbauer.}
\address{Northeastern University, 360 Huntington Ave, Boston Massachusetts 02118, USA.}

\begin{abstract}
  Dichotomy theorems, which characterize the conditions under which a problem can be solved efficiently, have helped identify important tractability borders for as probabilistic query evaluation, view maintenance, query containment (among many more problems).
  However, dichotomy theorems for many such problems remain elusive under key settings such as bag semantics or for queries with self-joins.
  This work aims to unearth dichotomies for fundamental problems in reverse data management and knowledge representation.
  We use a novel approach to discovering dichotomies: instead of creating dedicated algorithms for easy ($\PTIME$) and hard cases ($\NP$-complete), we devise \emph{unified algorithms} that are guaranteed to terminate in $\PTIME$ for easy cases.
  Using this approach, we discovered new tractable cases for the problem of minimal factorization of provenance formulas as well as dichotomies under bag semantics for the problems of resilience and causal responsibility.
\end{abstract}


\maketitle

\section{Introduction}

\begin{figure*}
  \centering
  \includegraphics[width=0.96\textwidth]{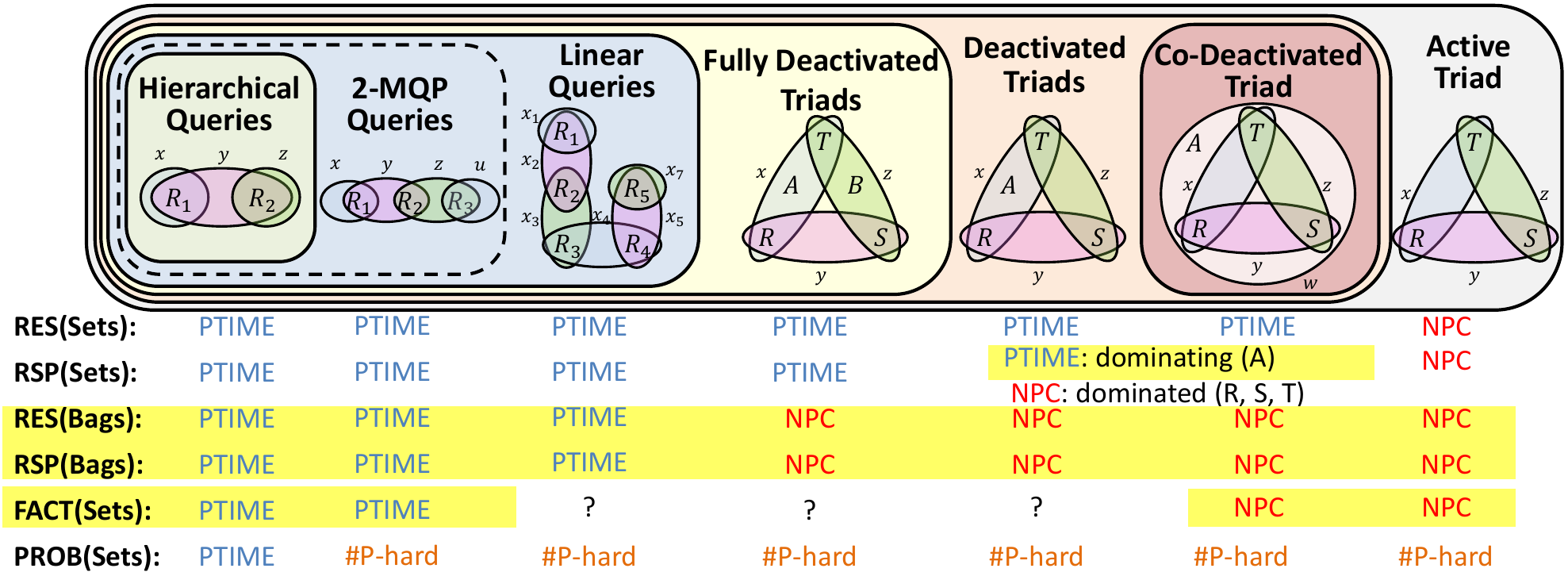}
  \captionof{figure}{Overview of complexity results for all self-join free conjunctive queries.
  The space of all queries is broken down into classes with different complexities as defined in the full papers~\cite{makhija2021minfac,makhija2022unified}.
  Results with a yellow background are new.
  $\res$ denotes resilience, $\rsp$ causal responsibility, $\fact$ minimal factorization, while $\prob$ denotes probabilistic query evaluation. 
\future{5/2: 
3. you have a bit more space horizontally 
5. is there a difference for PROB between sets and bags?}
  }
  \label{fig:queryvenn}
  \vspace{-15pt}
\end{figure*}%

Our goal is to understand the complexity of solving three distinct, yet related, problems: 
resilience, causal responsibility and minimal factorization. 
They underlie many practical problems such as reverse data management~\cite{DBLP:journals/pvldb/MeliouGS11}, (including view maintenance~\cite{Buneman:2002,Dayal82,ADP}, explanations and diagnostics~\cite{SudeepaSuciu14,glavic2021trends,wang2017qfix}), 
knowledge representation as boolean formulas~\cite{buchfuhrer2011complexity}, and probabilistic inference~\cite{DBLP:journals/vldb/DalviS07}. 
\wolf{5/2: maybe here in the intro just a bit more, like reverse data management, that it underly to many problems with a wide range of practical applications.
Then: We focus on the complexity.}

We treat all problems with the same novel method: 
Rather than deriving a dedicated $\PTIME$ algorithm for certain queries
(and proving hardness for the rest), 
we propose a unified Integer Linear Program (ILP) formulation for \emph{all} conjunctive queries under \emph{all} problem variants.
We then show that, for all $\PTIME$ queries, \emph{the Linear Program (LP) relaxation of our ILP has the same optimal value}, proving that existing ILP solvers are \emph{guaranteed} to solve problems for those queries in $\PTIME$.
Through this method, we are able to uncover new tractable cases and show the first dichotomy under bag semantics in this space~\cite{makhija2022unified} (\cref{fig:queryvenn}). 
In addition, we provide a unified hardness criterion by proving a prior conjecture~\cite{DBLP:conf/pods/FreireGIM20} that states that if a query can form an ``Independent Join Path'', then it must be hard~\cite{makhija2022unified}. 
This unified hardness criterion helps us prove the hardness of some queries with self-joins, whose complexities were previously unknown.

\introparagraph{Resilience}
What is the minimal number of tuples to delete from a database to eliminate all query answers? 

\begin{problem}[Resilience] 
How surprising is it for an Oscar winning actor to act in a movie directed by their spouse? 
We can quantify this by calculating the resilience of the query $\qtriangleunary \datarule$ Oscar(actor), ActsIn(actor, movie), DirectedBy(movie, dir), Spouse(actor, dir).
Resilience does not equate to simply the number of output rows but rather asks for the minimum number of changes in the world needed to have no satisfying output.
For example, if we do not include the spouse pair of Frances McDormand and Joel Coen, the single deletion would take away 9 rows from the output.
Intuitively, if the resilience is small, there have been a small number of events that have led to an Oscar winning actor being in a movie directed by their spouse.

Interestingly, the resilience for this query can be calculated in $\PTIME$ under set semantics, but not bag semantics (such as when accounting for multiple Oscar wins).
If we now change the query to remove the constraint of the actor having won an Oscar, then finding the resilience of the resulting query  $\qtriangle \datarule$ ActsIn(actor, movie), DirectedBy(movie, dir), Spouse(actor, dir) is $\NPC$!
\end{problem}

Resilience was introduced~\cite{DBLP:journals/pvldb/FreireGIM15} to capture the essence of all reverse data management questions: What is a minimum set of changes to a database to produce a certain change in the output of a query?
The same work gave a dichotomy~\cite{DBLP:journals/pvldb/FreireGIM15} of the complexity for resilience for self-join free queries under set semantics.
While a dichotomy for the general self-join case remains open, 
Freire et al.~\cite{DBLP:conf/pods/FreireGIM20} gave partial complexity results in this space, 
and conjectured that the notion of Independent Join Paths (IJPs) is a sufficient and necessary condition for hardness of resilience.
We proved this conjecture for self-join free queries (with a slight fix of the original statement)~\cite{makhija2022unified}, and proved that IJPs are a sufficient condition for hardness in queries with self-joins as well.

\introparagraph{Causal Responsibility}
For a given tuple in the output, what is a minimum subset of input tuples to remove to make the given tuple ``counterfactual''?

\begin{problem}[Causal Responsibility]
  Assume we wished to ask: ``What is the responsibility of Frances McDormand's Oscar win towards the output of our query?''
  If this Oscar was solely responsible for the output, it would be a counterfactual cause -- 
  i.e.\ if she had not won, 
  there would be no satisfying output.
  However, this tuple still has ``partial'' responsibility.
  By measuring how far we are from a world where the tuple is counterfactual, we can get a notion of its responsibility to the output.\footnote{The responsibility is inversely proportional to the minimum number of tuples to be deleted $|\tau|$ and is given by $1 / (1+|\tau|) $. }
  Interestingly, due to our new fine-grained complexity results, 
  we can find the responsibility of a particular Oscar win in $\PTIME$, but finding the responsibility of a tuple from 
  the ActsIn, DirectedBy or Spouse Table is $\NPC$.
\end{problem}

This notion of causal responsibility is due to foundational work by Halpern, Pearl, et al.~\cite{HalpernPearl:Cause2005}
which defined causal responsibility based on \emph{minimal interventions} in the causal graph. 
Meliou et al.~\cite{DBLP:journals/pvldb/MeliouGMS11} adapted this concept to define causal responsibility for database queries and showed a dichotomy under set semantics for self-join free queries.
Our work proves a dichotomy under bag semantics, as well as more fine-grained result based on the relation of the tuple we find responsibility of~\cite{makhija2022unified}.

\introparagraph{Minimal Factorization} 
Given the provenance formula for a Boolean query, what is its \emph{minimal size equivalent expression}?

\begin{problem}[Minimal Factorization]
  For the same query $\qtriangleunary$, how could we representing its output efficiently?
  Imagine the output was due to a single Oscar winning actor $o_1$, who was married once represented by $s_1$, and acted in two movies $a_1$, $a_2$, while their spouse also directed these same movies according to tuples $d_1$, $d_2$.
  We could then represent the output of this query with the boolean provenance formula $o_1 s_1 a_1 d_1 +$ $o_1 s_1 a_2 d_2$. However, notice that a factorized representation $o_1 s_1 (a_1 d_1 + a_2 d_2)$ is also equivalent (and minimal in this case). 
  We showed that the minimal factorization for $\qtriangleunary$ can be found in $\PTIME$~\cite{makhija2021minfac}, however the complexity of self-join free queries remains open in general.
\end{problem}

Minimal factorization solves the Minimum Equivalent Expression (MEE) 
~\cite{garey1979computers,buchfuhrer2011complexity} 
problem for provenance formulas.
Previously work~\cite{olteanu2012factorised} on this problem has lead to work on factorized databases~\cite{DBLP:journals/sigmod/OlteanuS16}.
However, the dichotomy for finding the exact minimal factorization is open.
Minimal factorizations of provenance can be used to obtain probabilistic inference bounds.
Prior approaches for approximate probabilistic inference are either incomplete i.e.\ focus on just $\PTIME$ cases~\cite{DBLP:journals/vldb/DalviS07,DBLP:conf/icdt/RoyPT11,SenDeshpandeGetoor2010:ReadOnce}, or do not solve all $\PTIME$ cases exactly~\cite{DBLP:journals/vldb/DalviS07,DBLP:journals/vldb/GatterbauerS17}. 
Using minimal factorization achieves the best of both worlds i.e. it applies to easy and hard cases while recovering all known $\PTIME$ cases exactly.
While we do not yet prove a dichotomy for minimal factorization, the quest for a dichotomy has helped prove a large tractable class: 
queries with \emph{$2$ Minimal Query Plans} ($2$-MQP).\footnote{We hypothesize that $2$-MQP Queries are a subset of linear queries.}
We also place the set of tractable queries for minimal factorization between resilience under set semantics and probabilistic inference in the self-join free case (\cref{fig:queryvenn})

\section{Unified ILP-Based Framework}

For all three problems, we can construct ILPs whose size (number of constraints and variables) is polynomial in the size of the database instance. 
While solving ILPs is $\NPC$ in general, we proved that the $\PTIME$ relaxations of our programs are correct for all easy cases, thus creating a unified framework that solves all known easy cases in $\PTIME$. 
We now provide intuition for these ILPs. For more details of construction and examples, please see the full papers~\cite{makhija2021minfac,makhija2022unified}.

\introparagraph{Resilience}
Intuitively, each tuple in the input is associated with a binary decision variable, which is set to $1$ if the tuple should be deleted else set to $0$.
The objective of the ILP is to minimize the number of tuples, which is the same as minimizing the sum of the variables.
For bag semantics, one can simply use a weighted sum, where the weight is the number of copies of a tuple in the database.
We have a constraint for each output row - that it must be deleted. 
An output row is deleted if any of the tuples contributing to it are deleted. 
Thus, for each output row, the sum of the variables of the tuple in the row must be $\geq 1$ i.e. at least one of the tuples must be deleted.

\introparagraph{Causal Responsibility}
To make a tuple counterfactual, one must delete all output rows it does not contribute to (which is identical to the resilience problem), but with the additional constraint that \emph{some output row must be preserved}. 
Thus, the causal responsibility ILP has the variables and constraints of the resilience ILP, plus additional variables to indicate if an output row is deleted, as well as one additional counterfactual constraint that enforces that all output rows cannot be deleted.

\introparagraph{Minimal Factorization}
The ILP for Minimal Factorization is based on the idea that different factorizations of a provenance formulas are equivalent to evaluating different output rows with different query plans. 
Thus, the ILP assigns a query plan to each output row.\footnote{We show that considering only the ``\emph{minimal query plans}'' suffices to obtain the minimal factorization.} 
The objective, equal to the length of the factorized expression, is calculated by summing up projections from the different query plans a tuple may be used in.

\section{Key Results}

\introparagraph{Result 1: ILP Relaxations recover all known $\PTIME$ cases}
We prove that for all prior known and newly found $\PTIME$ cases for all three problems, our ILPs are solved in guaranteed $\PTIME$ by standard solvers.
For the problems of resilience~\cite{makhija2022unified} and minimal factorization~\cite{makhija2021minfac}, we use the LP relaxation, i.e.\ all integrality constraints on all decision variables are removed. 
However, for causal responsibility, we relax all variables except those that track if an output row is preserved, creating a Mixed Integer Linear Program (MILP). While MILPs take exponential time in general, we show that the proposed MILP relaxation for causal responsibility can be solved in $\PTIME$ if the query is known to be in $\PTIME$~\cite{makhija2022unified}.

\introparagraph{Result 2: Dichotomies under Bag Semantics}
Real-world databases consist of bags instead of sets i.e. a relation may have multiple copies of the same tuple.
However, the complexity of the many problems under bag semantics is not well understood.
We show that both resilience and responsibility are easy under bag semantics if and only if the query is \emph{linear}~\cite{makhija2022unified}.
Notice that the tractability frontier for bag semantics differs from set semantics, where responsibility is a strictly harder problem than resilience (\cref{fig:queryvenn}). 
We show that switching from set semantics to bag semantics need only change the objective function of the ILP, and the constraint matrix remains the same.

\introparagraph{Result 3: New Tractable Cases}
  We are able to show that for any query with $\leq 2$ minimal query plans, minimal factorization can be solved in $\PTIME$ as the relaxation of the minimal factorization ILP is always correct~\cite{makhija2021minfac}. 
  This large class of queries includes hierarchical queries as a special case.
  In addition, a fine-grained analysis of the complexity of causal responsibility based on the relation of the tuple we find the responsibility of, reveals additional tractable cases (\cref{fig:queryvenn}).

\introparagraph{Result 4: Instance-Based Tractability}
  We prove cases such that our unified algorithm is \emph{guaranteed to terminate in $\PTIME$}, even for hard queries.
  For example, we show that when the provenance of a query output is read-once~\cite{SenDeshpandeGetoor2010:ReadOnce}, all three problems can be solved in $\PTIME$.
  The interesting aspect is that our unified algorithm \emph{does not need to know about these conditions as input}, it just automatically leverages those during query time.

\introparagraph{Result 5: Unified Hardness Criterion}
Freire et al.~\cite{DBLP:conf/pods/FreireGIM20} conjectured that the ability to construct Independent Join Paths (IJPs) is a necessary and sufficient criterion to prove hardness of resilience for a query.
We proved this conjecture for the self-join free case, and the sufficiency criterion for all cases\footnote{With a slight generalization of the definition of IJP~\cite{makhija2022unified}}, and use it as a unified way to show hardness for all three problems. 
With this unified hardness criterion, we obtain considerably simplified proofs, and prove hardness for a query with self-join whose complexity was previously unknown~\cite{makhija2022unified}. 
We give a Disjunctive Logic Program (DLP) formulation that can computationally derive such hardness certificates, and thus has shown several queries whose complexity was previously unknown to be hard.
We also show, with the existence of an IJP for query $A(a), R(a,x,y), S(a,y,z), T(a, z, x)$ that the tractable cases for minimal factorization are a strict subset of the tractable cases for resilience under set semantics.

\section{Conclusion and Future Work}

This overview gives the intuition for a novel way of determining the complexity of problems such as resilience, causal responsibility and minimal factorization.
We give a universal encoding as ILP and show that a relaxation is guaranteed to give an integral solution for all
$\PTIME$ cases, thereby proving that modern solvers can return the answer in guaranteed $\PTIME$.
While this approach is known in the optimization literature~\cite{schrijver2003combinatorial},
it has so far not been applied as \emph{proof method} to establish dichotomy results in reverse data management to the best of our knowledge.
Since the resulting theory is simpler and naturally captures all prior known $\PTIME$ cases, 
we believe that this approach will also help in related open problems in data management, in particular a so far elusive complete dichotomy for resilience of queries with self-joins~\cite{DBLP:conf/pods/FreireGIM20} and completing the dichotomy for minimal factorization.
The techniques in this paper are not limited to the three problems we consider, but may be helpful to discover other dichotomies as well.
\bibliography{BIB/vldbphdworkshop}

\end{document}

%% file: datalabmacros.tex
\usepackage{amsmath}
\usepackage{graphicx} %
\usepackage{url}
\usepackage{enumitem} %

\usepackage{bm} %

\usepackage{upgreek} %
\usepackage{soul} %
\usepackage{bbding} %
\usepackage{pifont} %

\usepackage{listings} %

\usepackage{tabularx} %
\usepackage{booktabs} %
\usepackage{multirow}
\usepackage{hhline} %
\usepackage{diagbox}
\usepackage{pgfplotstable} %

\usepackage{colortbl}
\definecolor{dg}{cmyk}{0.60,0,0.88,0.27}	

\usepackage{booktabs}  %

\usepackage{easybmat} %

\usepackage{rotating}		%

\usepackage{tikz} %
\usetikzlibrary{matrix}
\usetikzlibrary{calc}
\usetikzlibrary{math}
\usetikzlibrary{positioning,chains,fit,shapes,calc}
\definecolor{myblue}{RGB}{80,80,160}
\definecolor{mygreen}{RGB}{80,160,80}

\usetikzlibrary{arrows,shapes,trees,backgrounds,automata}
\usetikzlibrary{decorations.markings}
\usepackage{tikzsymbols}

\usepackage{subcaption} %
\usepackage[ruled,noend,linesnumbered]{algorithm2e} %

\DontPrintSemicolon     %
\SetNlSty{}{}{}                %
\SetAlgoInsideSkip{smallskip}   %
\SetAlCapSkip{1.5mm}             %
\SetInd{1.5mm}{1.5mm}             %

\SetAlFnt{\small}			%
\SetAlCapFnt{\small}		%
\SetAlCapNameFnt{\small}

\SetCommentSty{mycommfont}

\newtheorem{problem}{Problem}

\newtheorem{questionW}{Question}
\newtheorem{resultW}{Result}

{\end{itshape}
}
\newcommand{\todo}[1]{{{\color{red}{[#1]}}}} %
\newcommand{\note}[1]{{{\color{blue}{[#1]}}}} %

\newcommand{\wolf}[1]{{{\color{red}{[\textbf{WG}: #1]}}}}
\newcommand{\neha}[1]{{{\color{orange}{[\textbf{NM}: #1]}}}}
\newcommand{\nikos}[1]{{{\color{orange}{[\textbf{NT}: #1]}}}}
\newcommand{\emily}[1]{{{\color{red}{[\textbf{EC}: #1]}}}}

\newcommand{\deprecate}[1]{{{\color{gray}{[#1]}}}} %

\newcommand{\future}[1]{{{\color{blue}{[#1]}}}} 	%

\newcommand{\hide}[1]{} 		%
\newcommand{\mkclean}{
    \renewcommand{\note}{\hide}
	\renewcommand{\wolf}{\hide}
	\renewcommand{\neha}{\hide}
	\renewcommand{\nikos}{\hide}
	\renewcommand{\todo}{\hide}
    \renewcommand{\deprecate}{\hide}	
    \renewcommand{\emily}{\hide}	
	\renewcommand{\future}{\hide}		
}

\usepackage[capitalise,nameinlink]{cleveref} %

\crefformat{equation}{(#2#1#3)}						%
\crefrangeformat{equation}{(#3#1#4) to~(#5#2#6)}
\crefmultiformat{equation}{(#2#1#3)}%
{ and~(#2#1#3)}{, (#2#1#3)}{ and~(#2#1#3)}

\usepackage{tcolorbox}		%
\tcbuselibrary{breakable,skins}		%

\tcbset{examplestyle/.style={
		enhanced jigsaw,	%
		colback=blue!08,	%
		colframe=blue!08,	%
		arc=2mm,
		boxrule=1pt,		%
		left=1mm,
		right=1mm,
		left skip=0mm,  %
		right skip=0mm, %
		top=-1mm,		%
		bottom=1mm,		%
		breakable,		%
		parbox = false,		%
		before={\par\pagebreak[0]\vspace{1mm}\parindent=0pt},		
		after={\par\pagebreak[0]\vspace{1mm}\parindent=0pt},				
		bottomrule = 0mm,
		boxsep = 0mm,					%
		topsep at break=0pt,			%
		bottomsep at break=0pt,			%
		pad at break=0mm,
		pad before break=0mm,		
		pad after break=1mm,		
		bottomrule at break=0mm,
		toprule at break=0mm,		
		}}

\tcolorboxenvironment{example}{examplestyle}
\tcolorboxenvironment{problem}{examplestyle}

\tcbset{qrboxstyle/.style={
		enhanced jigsaw,	%
		colback=gray!20,
		colframe=gray!40,
		arc=0mm,
		boxrule=1pt,		%
		left=1pt,
		right=1pt,
		topsep at break=1mm,			%
		top=1pt,		%
		bottom=0mm,		%
		breakable,		%
		parbox = false		%
		}}

\newcounter{resultboxenv}

\newsavebox{\coloredbgbox}

\RequirePackage{scalerel}
\RequirePackage{tikz}
\usetikzlibrary{svg.path}

\definecolor{orcidlogocol}{HTML}{A6CE39}
\tikzset{
  orcidlogo/.pic={
    \fill[orcidlogocol] svg{M256,128c0,70.7-57.3,128-128,128C57.3,256,0,198.7,0,128C0,57.3,57.3,0,128,0C198.7,0,256,57.3,256,128z};
    \fill[white] svg{M86.3,186.2H70.9V79.1h15.4v48.4V186.2z}
                 svg{M108.9,79.1h41.6c39.6,0,57,28.3,57,53.6c0,27.5-21.5,53.6-56.8,53.6h-41.8V79.1z M124.3,172.4h24.5c34.9,0,42.9-26.5,42.9-39.7c0-21.5-13.7-39.7-43.7-39.7h-23.7V172.4z}
                 svg{M88.7,56.8c0,5.5-4.5,10.1-10.1,10.1c-5.6,0-10.1-4.6-10.1-10.1c0-5.6,4.5-10.1,10.1-10.1C84.2,46.7,88.7,51.3,88.7,56.8z};
  }
}

\RequirePackage{etoolbox}
\DeclareRobustCommand\orcidicon[1]{\href{https://orcid.org/#1}{\mbox{\scalerel*{
\begin{tikzpicture}[yscale=-1, transform shape]
    \pic{orcidlogo};
\end{tikzpicture}
}{|}}}}

\usepackage{adjustbox}
\def\eox{\unskip\kern 10pt{\unitlength1pt\linethickness{.4pt}$\diamondsuit${}}}

\makeatletter
\DeclareRobustCommand*\uell{\mathpalette\@uell\relax}
\newcommand*\@uell[2]{
  \setbox0=\hbox{$#1\ell$}
  \setbox1=\hbox{\rotatebox{10}{$#1\ell$}}
  \dimen0=\wd0 \advance\dimen0 by -\wd1 \divide\dimen0 by 2
  \mathord{\lower 0.1ex \hbox{\kern\dimen0\unhbox1\kern\dimen0}}
}
\makeatother

\newcommand{\introparagraph}[1]{\textbf{#1.}} %

\renewcommand{\epsilon}{\varepsilon} %

\newcommand{\datarule}{{\,:\!\!-\,}} %

\renewcommand{\P}{{\mathbb{P}}} %

\newcommand{\fact}{\mathtt{FACT}}

\newcommand{\res}{\mathtt{RES}} 
\newcommand{\rsp}{\mathtt{RSP}} 
\newcommand{\prob}{\mathtt{PROB}}

\newcommand{\PTIME}{\textup{\textsf{PTIME}}\xspace} 
\newcommand{\NPC}{\textup{\textsf{NPC}}\xspace}

\newcommand{\NP}{\textup{\textsf{NP}}\xspace}

\newcounter{myeqn}

\usepackage{xspace}            %

\renewcommand{\S}{\mathcal{S}}

\newcommand{\qtriangle}{Q^{\triangle}}
\newcommand{\qtriangleunary}{Q_{A}^{\triangle}}

\definecolor{dg}{cmyk}{0.60,0,0.88,0.27}

\usepackage{graphbox} %

\usetikzlibrary{tikzmark, calc}
\tikzset{
    double color fill/.code 2 args={
        \pgfdeclareverticalshading[%
            tikz@axis@top,tikz@axis@middle,tikz@axis@bottom%
        ]{diagonalfill}{100bp}{%
            color(0bp)=(tikz@axis@bottom);
            color(50bp)=(tikz@axis@bottom);
            color(50bp)=(tikz@axis@middle);
            color(50bp)=(tikz@axis@top);
            color(100bp)=(tikz@axis@top)
        }
        \tikzset{shade, left color=#1, right color=#2, shading=diagonalfill}
    }
}
\newcounter{BGnum}
\usepackage{dcolumn, array, booktabs}
\newcolumntype{.}{D{.}{.}{-1}}

\usepackage{thmtools}
\usepackage{thm-restate}

\setlist[itemize,1]{leftmargin=\dimexpr 10pt}
\setlist[enumerate,1]{leftmargin=\dimexpr 15pt}

\usepackage{listings}